\documentclass[12pt]{iopart}

\usepackage{graphicx}   
\begin{document}
\title[Reply to comment on `Poynting flux ... and the radiative power losses']
{Reply to comment on `Poynting flux in the neighbourhood of a point charge in arbitrary motion and the radiative power losses'} 
\author{Ashok K. Singal}
\address{Astronomy and Astrophysics Division, Physical Research Laboratory,
Navrangpura, Ahmedabad - 380 009, India }
\ead{ashokkumar.singal@gmail.com}
\vspace{10pt}
\begin{indented}
\item[]August 2017
\end{indented}
\begin{abstract}
Doubts have been expressed in a comment \cite{row18}, about the tenability of the formulation for radiative losses in our recent published work \cite{68a}. We provide our reply to the comment. In particular, it is pointed out that one need to clearly distinguish between the rate of the energy-momentum being carried by the electromagnetic radiation to far-off space, and that of the mechanical energy-momentum losses being incurred by the radiating charge. It is also demonstrated that while the Poynting flux is always positive through a spherical surface centred on the retarded position of the charge, it could surprisingly be negative through a surface centred on the ``present'' position of the charge. It is further shown that the mysterious Schott term, hitherto thought in literature to arise from some acceleration-dependent energy in fields, is actually nothing but the difference in rate of change of energy in self-fields of the charge between the retarded and present times.
\end{abstract}
This is a reply to comment by Rowland \cite{row18} on my recently published work \cite{68a}. All points raised by Rowland are discussed below, though not in same order. Like Rowland, we shall also confine our discussion largely to non-relativistic motion, unless otherwise specified.
\section{A difference in the physical interpretation of two power formulas}
We begin by pointing out the difference in the physical interpretation of the two power formulas in question. 
First is the well-known Larmor's formula, representing the power going into electromagnetic radiation from an accelerated charge 
\begin{equation}
\label{eq:6.2}
{\cal P}_1 =\frac{2e^{2}}{3c^{3}}[\dot{\bf v}^2]_{\rm ret}\:,
\end{equation}
This formula has a text-book derivation \cite{1,2,25} where Poynting flux through a spherical surface of large enough radius is computed from acceleration fields, assuming any contribution of velocity fields could be neglected. 
The latter condition is met in almost all cases, a notable exception being where the velocity of the accelerated charge 
may be a monotonic function of time, e.g., in the case of a uniformly accelerated charge \cite{18}. 
Leaving apart any such peculiar cases, Larmor's formula does represent electromagnetic radiation power which will be received by a set of distant observers stationed on a 
spherical surface of radius $r$ around the retarded position of the charge. Thereafter, in the standard text-book approach, one equates  the Poynting flux at time $t$ to the kinetic energy loss rate of the 
charge at a retarded time $t-r/c$, purportedly using Poynting's theorem of energy conservation. 
However, a fallacy lies in this particular step. Poynting's theorem does 
not relate Poynting flux through a surface at some time $t$ to energy loss rate by the enclosed 
charge at a retarded time $t-r/c$. In fact most of the confusion in this hundred years old controversy has arisen due to this oversight. In Poynting's theorem all quantities need to be calculated for the {\em same instant of time} \cite{1,2,25}. Applying Poynting's theorem correctly in terms of real time values of the charge motion \cite{68a}, one gets the 
instantaneous rate of loss of the mechanical energy by the charge as 
\begin{equation}
\label{eq:6.1}
{\cal P}_2 =-\frac{2e^{2}}{3c^{3}}\ddot{\bf v}\cdot{\bf v}\:.
\end{equation}

One should clearly distinguish between the electromagnetic power received 
by a set of far-off observers and the instantaneous loss of mechanical power by the charge. In literature both power rates are treated as not only equal but almost synonymous. However, the two need not be the same as seen from Eqs.~(\ref{eq:6.2}) and (\ref{eq:6.1}). The difference in the two power formulas is 
\begin{equation}
\label{eq:9.1}
{\cal P}_2-{\cal P}_1=-\frac{2e^{2}}{3c^{3}}\ddot{\bf v}\cdot{\bf v} -\frac{2e^{2}}{3c^{3}}\dot{\bf v}\cdot\dot{\bf v}
=-\frac{2e^{2}}{3c^{3}}\frac{{\rm d}(\dot{\bf v}\cdot{\bf v})}{{\rm d}t}\:.
\end{equation}
The last term on the right hand side in Eq.~(\ref{eq:9.1}) is known as the Schott term, after Schott \cite{7} who first pointed it out, and is thought in literature to arise from an acceleration-dependent energy in electromagnetic fields. 
The meaning of this elusive, century-old term is still being debated \cite{6,44,41,56,row12} 
and it does not seem to make an appearance elsewhere in physics. We shall later demonstrate that the Schott term arises as a consequence of not keeping a proper distinction between ``real'' and ``retarded'' times while calculating power losses for a radiating charge. 

The rate of momentum being carried away by the electromagnetic radiation (due to its $\sin^2\theta$ pattern) in the instantaneous rest frame of the charge is zero. However in the frame where the charge moves with a velocity ${\bf v}$, one gets \cite{51,ha95}
\begin{equation}
\label{eq:21b}
\dot{\bf p}_{1}=\frac{{\cal P}_{1}}{c^2}{{\bf v}}\:.
\end{equation}
Now contrary to the view expressed by Rowland in his comment (his Eq. (7) and the discussion following that), the negative of $\dot{\bf p}_{1}$ cannot be the radiation reaction on the charge, as an application of Eq.~(\ref{eq:21b}) along with Eq.~(\ref{eq:6.2}) to a radiating synchrotron source case leads to results which are mutually inconsistent when compared in two different inertial reference frames \cite{68a}.

On the other hand, the momentum conservation theorem, using Maxwell's stress tensor, directly leads to a rate of change of momentum of the charge \cite{68b} 
\begin{equation}
\label{eq:3a}
\dot{\bf p}_{2}= \frac{2e^{2}}{3c^{3}}\ddot{\bf v}\:.
\end{equation}
The result in  Eq.~(\ref{eq:3a}), known as the Abraham-Lorentz radiation reaction formula, has been obtained earlier from the self-force of the charge, calculated albeit in a rather cumbersome manner \cite{1,2,7,16,20}.

It should be noted that Eq.~(\ref{eq:6.1}), representing the rate of change in the mechanical energy of a charge, was not derived using the radiation drag force (Eq.~(\ref{eq:3a})) as mentioned in the comment by Rowland after his Eq. (8), but was instead calculated directly from Poynting flux when written in terms of real time values of the charge motion \cite{68a}. The result though does turn out to be consistent with the work being done against the drag force $\dot{\bf p}_{2}$ (Eq.~(\ref{eq:3a})).

\section{Absence of radiation from a uniformly accelerated charge}
Using the vector identity ${\bf v}={\bf n}({\bf v}.{\bf n}) - {\bf n}\times\{{\bf n}\times{\bf v}\}$, the 
transverse component of the electric field of an accelerated charge, moving with a non-relativistic velocity ${\bf v}$ and an acceleration $\dot{\bf v}$ at the retarded time, can be written as
\begin{equation}
\label{eq:1a}
{\bf E}_T=\frac{e{\bf n}\times({\bf n}\times {\bf v})} {cr^{2}} + 
\frac{e{\bf n}\times({\bf n}\times \dot{\bf v})}{c^2r}.
\end{equation}
The conventional wisdom is that the acceleration fields ($\propto 1/r$) solely represent the radiation from a charge, 
since the contribution of the velocity fields ($\propto 1/r^{2}$) appears to be negligible for a large enough value of $r$. 
However, in the case of a uniform 
acceleration, the {\em retarded value} of the velocity will be ${\bf v} = {\bf v}_0 -\dot{\bf v} r/c$, where ${\bf v}_0$ is the {\em present} velocity of the charge. Then Eq.~(\ref{eq:1a}) for the transverse component of the electric field becomes 
\begin{eqnarray}
\label{eq:1a11}
\!\!\!\!\!\!\!\!{\bf E}_{T}=\frac{e{\bf n}\times({\bf n}\times {\bf v_0})} {cr^{2}} - \frac{e{\bf n}\times({\bf n}\times \dot{\bf v})r}{c^2r^2}
+\frac{e{\bf n}\times({\bf n}\times \dot{\bf v})}{c^2r}=\frac{e{\bf n}\times({\bf n}\times {\bf v}_0)} {cr^{2}},
\end{eqnarray}
which for the Poynting flux begets 
\begin{equation}
\label{eq:11c1}
{\cal P}= \frac{2e^{2}}{3r^2c} {\bf v}_0^{2}.
\end{equation}

The transverse component of the electric field here (Eq.~(\ref{eq:1a11})) is the same as would be that of a charge moving with a uniform velocity ${{\bf v}}_{\rm o}$, equal to the ``present'' velocity of the accelerated charge. It is clear that in the case of a uniformly accelerated charge, in the Poynting flux expression (Eq.~(\ref{eq:11c1})), there is no term proportional to $\dot{\bf v}^{2}$, 
independent of $r$, which is commonly called the radiated power. Nor is there any term proportional to ${\bf v}\cdot\dot{\bf v}$, the Schott energy term, as suggested in Eq. (10) of Rowland. Instead, the Poynting flux in Eq.~(\ref{eq:11c1}) is merely what would be for a charge moving with a uniform velocity ${{\bf v}}_{\rm o}$, justifying its usage in Eq. (17) of \cite{68a}. 

\section{No power losses from an instantaneously stationary charge}
It is already evident from Eqs.~(\ref{eq:1a11}) that 
the transverse component of electric field in the instantaneous rest frame (${{\bf v}}_{\rm o}=0$) of a uniformly accelerated charge is nil, as the acceleration fields there get cancelled {\em neatly} by the transverse term of the velocity fields {\em at all distances}. Consequently there is a nil Poynting flux through any surface around such a  charge (Eqs.~(\ref{eq:11c1})),
which is consistent with the assertion made in \cite{68a} that there is no radiation from an accelerated charge that is instantaneously stationary. 

Now from Poynting's theorem we shall explicitly demonstrate that there are no radiative losses from a uniformly accelerated charge in its instantaneous rest-frame. 
Let a charge moving with a uniform acceleration $a$ along the $+z$ axis, starting from $z=-\infty$ at time $t=-\infty$, momentarily comes to rest at a point $z=\eta$ at time $t=0$, and 
then onwards moves with an increasing velocity along the $+z$ axis.
Choosing the origin of the coordinate system so that $\eta=c^{2}/a$,
the position and velocity of the charge at any time $t$ are given by 
$z_{0}=(\eta^{2}+c^{2}t^{2})^{1/2}$ and $v=c^{2}t/z_{0}$.
The electromagnetic fields in cylindrical coordinates ($\rho,\phi ,z$)
at any instant $t$ are given by \cite{5} 
\begin{eqnarray}
\label{eq:32a1}
E_{z}&=&-4e\eta^{2}(z_{0}^{2}-z^{2}+\rho^{2})/\xi^{3}
\nonumber\\
E_{\rho}&=&8e\eta^{2}\rho z/\xi^{3}\nonumber\\
B_{\phi}&=&8e\eta^{2}\rho ct/\xi^{3}\;,
\end{eqnarray}
where $\xi=[(z_{0}^{2}-z^{2}-\rho^{2})^2+4\eta^{2}\rho^{2}]^{1/2}$. All other field components are zero. 
We shall restrict all discussions to a region $z+ct>0$ 
because it is only within this region that the light signals
from the retarded positions of the charge could have reached \cite{5}. 

The first thing we notice from Eq.~(\ref{eq:32a1}) is that at time $t=0$, when the charge has come to rest momentarily at 
$z_{0}=\eta$, the magnetic field is zero throughout. Therefore the Poynting flux from any closed surface $\Sigma$ surrounding the charge will be zero. 
\begin{equation}
\label{eq:32a2}
\int_{\Sigma}{{\rm d}\Sigma}\:({\bf n} \cdot {\cal S})=0 \;.
\end{equation}
Further, the field energy density, $({E^{2}+B^{2}})/{8\pi}$, equal at times $t$ and $-t$, is an even function of $t$. 
Therefore the electromagnetic field energy in the  volume enclosed by  $\Sigma$ 
\begin{equation}
\label{eq:32a3}
{\cal E}_{\rm em}=\int \frac{E^{2}+B^{2}}{8\pi}\:dv \;,
\end{equation}
is also an even function of $t$, implying 
\begin{equation}
\label{eq:32a4}
\frac{{\rm d}{\cal E_{\rm em}}}{{\rm d}t}=0,
\end{equation}
at $t=0$. Then from the Poynting's theorem, the rate of change of the mechanical energy $({\cal E_{\rm me}})$ of the 
charge at $t=0$ given by
\begin{equation}
\label{eq:21a}
\frac{{\rm d}{\cal E_{\rm me}}}{{\rm d}t}=-\frac{{\rm d}{\cal E_{\rm em}}}{{\rm d}t}-\int_{\Sigma}{{\rm d}\Sigma}\:({\bf n} \cdot {\cal S})=0 \;,
\end{equation}
where all quantities are evaluated at the same time, $t=0$. This immediately implies no power losses by the instantly stationary charge.
Here we see no sign of the acceleration-dependent Schott term ($\propto {\rm d}({\bf v\cdot \dot{v}})/{{\rm d}t}$, mentioned in the comment by Rowland) which is supposed to make the net power loss from a uniformly accelerated charge zero even when there is radiated power as per Larmor's formula (the latter is also not seen here).

\section{Schott energy term -- a difference in the self-field  energy of an accelerated charge between retarded and present time}
In order to better apprehend the difference between Eqs.~(\ref{eq:6.2}) and (\ref{eq:6.1}) and their relations with  retarded and real times, we consider the effect of the self-force of an accelerated charge on itself. For this we take the charge to be a spherical shell of a small radius $\epsilon$, in order to avoid divergence of fields at the centre of a point charge, though the final results turn out to be independent of the radius assumed for the sphere. Force 
on each infinitesimal element of the spherical shell is calculated 
due to the time-retarded fields from the remainder parts 
of the charged shell and then total force on the charge is calculated by integrating over the whole shell.

It has been shown \cite{68} that for an accelerated charge, there is a net self-force proportional to the acceleration which is as if due to the time-retarded fields of a co-moving, equivalent point charge at the centre \cite{50}. Thus the charged spherical shell experiences a force proportional to the acceleration it 
had at a time interval $\tau=\epsilon /c$ earlier, because of fields from the centre delayed due to the finite speed $c$ of propagation.
Effectively the self-force on the charge at time $t$ is then proportional to the acceleration it had at a retarded time $t_{\rm o}=t-\epsilon /c$
\begin{equation}
\label{eq:3d1}
{\bf f}_t=-\frac{2e^{2}}{3\epsilon c^{2}}[\dot{\bf v}]_{t_{\rm o}}\:,
\end{equation}
where a square bracket denotes a retarded-time value \cite{68}.
Accordingly, for an accelerated charge, the power loss during work done against self-force of the charge is given by 
\begin{equation}
\label{eq:6}
{\cal P}_t=-{\bf f}_t \cdot {\bf v}_t=\frac{2e^{2}}{3\epsilon c^{2}}[\dot{\bf v}]_{t_{\rm o}} \cdot {\bf v}_t\:.
\end{equation}
Now if we write the velocity too in terms of its 
value at the retarded time ${t_{\rm o}}$ (to a first order in $\epsilon/c$) 
\begin{equation}
\label{eq:21.1}
{\bf v}_t=[{\bf v}]_{t_{\rm o}}+[\dot{\bf v}]_{t_{\rm o}}\:{\epsilon}/{c}
\end{equation}
we get  
\begin{equation}
\label{eq:9}
{\cal P}_t=\frac{2e^{2}}{3\epsilon c^{2}}[\dot{\bf v}\cdot
{\bf v}]_{t_{\rm o}}+\frac{2e^{2}}{3c^{3}}[\dot{\bf v}\cdot\dot{\bf v}]_{t_{\rm o}} \:.
\end{equation}
On the right hand side, the first term shows the rate of change of self-field energy of the accelerated charge due to its changing momentum {\em at the retarded time}, and the second term comprises Larmor's formula, again evaluated at the retarded time.

However if we instead express the acceleration itself in terms of its  real-time value at $t={t_{\rm o}}+\tau$ 
\begin{equation}
\label{eq:22}
[\dot{\bf v}]_{t_{\rm o}} =\dot{\bf v}_t-{\ddot{\bf v}_t}\tau+\cdots, 
\end{equation}
then the self-force (Eq.~(\ref{eq:3d1})) can be written in terms of real-time values as 
\begin{equation}
\label{eq:3a1}
{\bf f}_t= -\frac{2e^{2}}{3\epsilon c^{2}}\dot{\bf v}_t+\frac{2e^{2}}{3c^{3}}\ddot{\bf v}_t\:,
\end{equation}
and the corresponding formula for power (Eq.~(\ref{eq:6})), in terms of real time, is  written as
\begin{equation}
\label{eq:6a}
{\cal P}_t=\frac{2e^{2}}{3\epsilon c^{2}}(\dot{\bf v}\cdot{\bf v})_t
-\frac{2e^{2}}{3c^{3}}(\ddot{\bf v}\cdot{\bf v})_t\:.
\end{equation}

Now the rate of change of energy in self-fields between retarded and real times (the leading terms on the right hand sides of Eqs.~(\ref{eq:9}) and (\ref{eq:6a})), differs by 
\begin{eqnarray}
\label{eq:9.2}
\frac{2e^{2}}{3\epsilon c^{2}}[\dot{\bf v}\cdot {\bf v}]_{t_{\rm o}}-\frac{2e^{2}}{3\epsilon c^{2}}(\dot{\bf v}\cdot{\bf v})_{t}
=-\frac{2e^{2}}{3\epsilon c^{2}}\frac{{\rm d}(\dot{\bf v}\cdot{\bf v})}{{\rm d}t}\tau
=-\frac{2e^{2}}{3c^{3}}\frac{{\rm d}(\dot{\bf v}\cdot{\bf v})}{{\rm d}t}\:,
\end{eqnarray}
a result independent of $\epsilon$. This explains the genesis of the mysterious Schott term and makes it obvious that this term, hitherto thought to arise from some acceleration-dependent energy in fields, is actually nothing but the difference in rate of change of energy in self-fields of the charge between retarded and  present times.
\section{Energy being radiated at a negative rate?}
Rowland in his comment has shown through a specific example, where acceleration of a charge is increasing in its direction of motion, Eq.~(\ref{eq:6.1}) yields a negative value, apparently implying energy being radiated at a negative rate. 
Actually in such cases it is not that the Poynting flux is negative (or inward) but that the rate of change of the kinetic energy of the charge is more than what should be expected from the Poynting flux calculated from its motion at the retarded time. We can elucidate it in the following manner.

From (Eq.~(\ref{eq:1a})), we can write the transverse component of the electric field as
\begin{equation}
\label{eq:1a1}
{\bf E}_T=\frac{e{\bf n}\times[{\bf n}\times ({\bf v}+\dot{\bf v}r/c)]} {cr^{2}},
\end{equation}
with ${\bf v}$ and ${\dot{\bf v}}$ being the (non-relativistic) velocity and acceleration at the retarded time. 
Then for the Poynting flux we get 
\begin{equation}
\label{eq:11c2}
{\cal P}= \frac{2e^{2}}{3r^2c} ({\bf v}+\dot{\bf v}r/c)^{2}.
\end{equation}
The expression $({\bf v}+\dot{\bf v}r/c)$ represents the extrapolated ``present'' velocity ${\bf v}_{0e}$ of the charge, assuming 
the acceleration $\dot{\bf v}$ remained constant during the time interval $\tau=r/c$.
The Poynting flux through a spherical surface centered on the retarded position of the charge is always given in accordance with (\ref{eq:11c2}), i.e.,  $\propto {\bf v}^2_{0e}/r ^2$. 
This is true for all $r$, including in the neighbourhood of the charge.

On the other hand, the Poynting flux through a spherical surface in the neighbourhood of the charge is also written as  \cite{68a}
\begin{equation}
\label{eq:1o2}
{\cal P}= \frac{2e^2({\bf v}_{\rm o}-\ddot{\bf v}_{\rm o}r^2/2c^2)^2}{3r^{2}c}= \frac{2e^2{\bf v}_{\rm o}^2}{3r^{2}c}-\frac{2e^2{\bf v}_{\rm o}\cdot\ddot{\bf v}_{\rm o}}{3c^{3}}\;.
\end{equation}

A comparison of Eqs.~(\ref{eq:11c2}) and (\ref{eq:1o2}) shows that ${\cal P}_2$ (Eqs.~(\ref{eq:6.1})) will be a negative quantity when ${\bf v}^2_{\rm o}>{\bf v}_{0e}^2$, as in the example by Rowland. On the other hand  ${\cal P}_2>0$ when 
${\bf v}^2_{\rm o}<{\bf v}_{0e}^2$, as happens in vast majority of cases, for instance in a circular motion or in an oscillatory motion where displace of charge from its equilibrium position gives rise to a restoring force proportional to the displacement. In no case is the Poynting flux through the spherical surface centered on the retarded position of the charge negative or inward (c.f. Eqs.~(\ref{eq:11c2}) and (\ref{eq:1o2})).

It may be worth pointing out here that the Poynting flux through the spherical surface centered on the ``present'' position of the charge could surprisingly be negative even though it is always positive through the surface centered on the retarded position of the charge. 
For example, in the case of a uniform acceleration, there 
is always an outward flow of Poynting flux through a spherical surface centered on the retarded position of the charge 
(Eq.~(\ref{eq:11c1})). However, in our chosen coordinate system in Section 3, the charge
occupies the same position $z_{0}$ at times $-t$ and $t$. 
We can choose a fixed finite closed surface, say, a sphere centered at $z_{0}$.
The Poynting vector, at any point 
on the surface, at time $t$ is exactly equal in magnitude but opposite in direction to its value at time $-t$. This  
is easily seen from Eq.~(\ref{eq:32a1}),
where the electric field vector is independent of sign of $t$ while the magnetic
field changes sign with $t$. Thus if there is an outward flow of Poynting
flux through the surface
at time $t$, it immediately follows that there was an equal 
inward flow of Poynting flux through that surface at time $-t$. 
Of course an even simpler example is that of a charge moving with a uniform velocity ${\bf v}_{\rm o}$, where there 
is always an outward flow of Poynting flux through a spherical surface centered on the retarded position of the uniformly moving charge 
(Eq.~(\ref{eq:11c1})), but at the same time we know that the electric field of a charge moving with a uniform velocity is 
radial with respect to its ``present'' position and therefore there is a nil Poynting flux through a surface centered on the present position of the charge. 
{}
\end{document}